\documentclass[%
 aip,
 jmp,
 amsmath,amssymb,
 reprint,%
]{revtex4-1}
\usepackage{mathptmx}
\usepackage{mathrsfs}
\usepackage{graphicx}
\usepackage{dcolumn}
\usepackage{bm}
\usepackage{caption}
\usepackage{CJK}
\usepackage{etoolbox}
\usepackage{float}
\usepackage{amstext}
\makeatletter
\def\@email#1#2{%
 \endgroup
 \patchcmd{\titleblock@produce}
  {\frontmatter@RRAPformat}
  {\frontmatter@RRAPformat{\produce@RRAP{*#1\href{mailto:#2}{#2}}}\frontmatter@RRAPformat}
  {}{}
}%
\makeatother
\begin{document}


\title[]{A single-particle energy-conserving dissipative particle dynamics approach for simulating thermophoresis of nanoparticles in polymer networks}
\author{Yu Lu
\begin{CJK}{UTF8}{gbsn}
(陆钰)
\end{CJK}
}
\affiliation{ 
School of Mechanical Engineering,
Nantong University,
Nantong 226002, China
}
\affiliation{ 
Shanghai Institute of Applied Mathematics and Mechanics, School of Mechanics and Engineering Science, Shanghai Key Laboratory of Mechanics in Energy Engineering,
Shanghai Frontier Science Center of Mechanoinformatics, Shanghai University,
Shanghai 200072, China
}
\author{Guo-Hui Hu
\begin{CJK}{UTF8}{gbsn}
(胡国辉)
\end{CJK}
$^*$}%
\email{ghhu@staff.shu.edu.cn}
\affiliation{ 
Shanghai Institute of Applied Mathematics and Mechanics, School of Mechanics and Engineering Science, Shanghai Key Laboratory of Mechanics in Energy Engineering,
Shanghai Frontier Science Center of Mechanoinformatics, Shanghai University,
Shanghai 200072, China
}
\affiliation{ 
Shanghai Institute of Aircraft Mechanics and Control, 100 Zhangwu Road, Shanghai 200092, China
}

\date{May 12, 2023}

\begin{abstract}
Thermophoresis is an effective method to drive the motion of nanoparticles in fluids. The transport of nanoparticles in polymer networks has significant fundamental and applied importance in biology and medicine, and can be described as Brownian particles crossing entropic barriers. This study proposes a novel extension of dissipative particle dynamics (DPD), called the single-particle energy-conserving dissipative particle dynamics (seDPD), which combines the features of single-particle dissipative particle dynamics (sDPD) and energy-conserving dissipative particle dynamics (eDPD) to simulate the thermophoresis of nanoparticles under temperature gradients. The reliability of the seDPD method is verified by considering the viscosity, thermal diffusivity, and hydrodynamic drag force on the nanoparticles. Using this method, the transport of nanoparticles driven by the thermophoretic force across the polymer network is simulated. The results show that the nanoparticles exhibit the phenomenon of giant acceleration of diffusion (GAD) in the polymer network, indicating that Brownian particles can exhibit GAD when crossing entropic barriers.
\end{abstract}

\maketitle
\section{introduction}

Thermophoresis\cite{huang2010isotope,wienken2010protein,dominguez2011soret,Duhr2006Thermophoretic,Reichl2014Why} is the phenomenon of nanoparticle migration under a temperature gradient, often also referred to as the Soret effect. It is an effective method to move nanoparticles in fluids.
Considering the size of nanoparticles ranging from tens to hundreds of nanometers\cite{wang2009supramolecular,xue2016probing,Xue2020Diffusion}, mesoscopic molecular dynamics methods are powerful tools for studying nanoparticle thermophoresis. One such mesoscopic method is dissipative particle dynamics proposed by Hoogerbrugge and Koelman\cite{hoogerbrugge1992simulating}, which has been applied to various systems such as DNA\cite{Lu2023Linear,Ranjith2014Transport}, rheology\cite{Boek1997Simulating,Fedosov2010Steady,Pasquino2019rheological}, drug release\cite{feng2020dissipative,wang2021dissipative}, and anomalous diffusion\cite{Lu2021potential,Lu2022Double,Xu2021Enhanced}. The standard DPD method follows Newton's second law, where the particles experience conservative, dissipative, and random forces. 
The conservative force is a soft repulsive force that ensures the compressibility of the fluid. The dissipative force acts as the fluid viscosity, and the random force represents the stochastic effects at the mesoscopic scale. The dissipative force and the random force obey the fluctuation-dissipation theorem.

The DPD method is a flexible and extensible technique for mesoscopic simulations. Several extensions of the DPD method have been proposed to model different physical phenomena. For example, the many-body dissipative particle dynamics (mDPD)\cite{Li2013Three} method adds an attractive term to the conservative force to simulate vapor-liquid coexistence systems. The energy-conserving dissipative particle dynamics (eDPD)\cite{Li2014Energy} method ensures the conservation of kinetic and internal energy by introducing temperature and heat flux in DPD, which enables the study of heat transfer and thermal convection. The single-particle dissipative particle dynamics (sDPD)\cite{Pan2008Single} method introduces a non-central dissipative force and solves the conservation of the angular momentum of the DPD particle, which allows the simulation of spherical nanoparticles with finite size by a single DPD particle. Furthermore, these extended DPD methods can be integrated with each other. For instance, Zhang {\it et al.}\cite{Zhang2022eDPDm} proposed a hybrid method of mDPD and eDPD, named mDPDe, which can simulate the thermocapillary motion of droplets on solid surfaces.

Despite the long history and versatility of the dissipative particle dynamics (DPD) method for simulating various systems, few existing method can model the thermophoretic behavior of nanoparticles using a single DPD particle. To address this gap, we propose a novel method that combines the sDPD method for nanoparticle simulation with the eDPD model for temperature and heat flux. Our method, which we call single-particle energy-conserving dissipative particle dynamics (seDPD). There are two technical issues for implementing the thermophoresis of nanoparticles in the seDPD method: (1) the viscous heat flux in eDPD needs to be modified due to the work done by the non-central dissipative and random forces in sDPD; (2) how to apply the thermophoretic force to the nanoparticles in the fluid with a temperature gradient.

We modify the conservative force in seDPD to account for the thermophoretic force $F^T$ that the temperature gradient $\nabla\Theta$ generates, which enables the nanoparticles to move along or against the gradient.
The Soret force $F^T$ depends on both the temperature gradient $\nabla\Theta$ and the particle size $R$.
Some previous studies have established the relation between $F^T$ and the size of nanoparticles as well as the temperature gradient.
Duhr {\it et al.}\cite{Duhr2006Thermophoretic} report experimental results for particles in liquids with temperature gradients, showing that the Soret coefficient $S_T=F_T/\nabla\Theta$ is proportional to the square of nanoparticle size ($S_T\sim R^2$). Mayer {\it et al.}\cite{Mayer2023Thermophoresis} measure the thermophoresis of polystyrene beads over a wide range of temperature gradients.
They find that the particle motion is dominated by fluctuations when the Peclet number $P_e =RS_T\nabla \Theta <1$, and that the Soret force varies linearly with the temperature gradient.
For large Peclet numbers ($P_e =RS_T\nabla \Theta > 1$), the Soret force varies sublinearly with the temperature gradient, the thermophoresis is dominated by drift and approaches the hydrodynamic model\cite{Rasuli2008Soret,Mayer2023Thermophoretic}. The results of these studies will help us to implement the correct thermophoresis in the seDPD method, and provide theoretical and experimental basis for the parameterization in seDPD.

In this work, we propose a modified DPD method, seDPD, to model the thermophoresis of nanoparticles in fluids. This method combines the sDPD with the eDPD. 
The validation of viscosity, heat conduction and hydrodynamic drag force on nanoparticles in seDPD is applied.
Furthermore, we develop a parameterization to control thermophoretic force on nanoparticles. Finally, we simulate the nanoparticle transport in the polymer network immersed with fluid with temperature gradients. This proposed model is not limited to thermophoresis of nanoparticles and can readily be extended to investigate various system including temperature and nanoparticles.

The paper is organized as follows. In section \ref{sec2}, we introduce the governing equations and parameters of seDPD, the strategy of choosing physical units, and the methods of calculating the thermophoretic force. In section \ref{sec3}, we validate the seDPD method by some benchmark cases to ensure its reliability in simulating the thermophoresis of nanoparticles, then we present the relation between the thermophoretic force and the seDPD parameters, and finally we simulate the transport of nanoparticles driven by the thermophoretic force in a polymer network. Finally, we conclude with a brief summary in section \ref{sec4}.

\section{model and method}\label{sec2}
\subsection{Governing equations of seDPD}
%
In seDPD method, the simulation system contains a collection of finite-size particles with mass $m_{i}$ and mass moment of inertia $I_{i}$ and heat capacity $C_{vi}$ for the $i$th particle.
The governing equation of particle motion is the conservation of momentum and energy:
\begin{equation}\label{equ:NSL}
\frac{d \vec{r}_{i}}{d t}=\vec{v}_{i} ; \ \ \  m_{i} \frac{d^{2} \vec{r}_{i}}{d t^{2}}=\vec{F}_{i}=\sum_{j} \vec{F}_{i j},
\end{equation}
\begin{equation}\label{equ:AM}
I_i\frac{d{\vec{\omega}}_i}{dt}={\vec{T}}_i=-\sum_{j}{\lambda_{ij}{\vec{r}}_i\times{\vec{F}}_{ij}},
\end{equation}
\begin{equation}\label{equ:Energy}
    C_{vi}\frac{\partial \Theta_i}{\partial t}=q_i=\sum_{i \neq j}q_{ij},
\end{equation}
where $\vec{r}_{i}, \vec{v}_{i}$ , $\vec{\omega}_{i}$ and $\Theta_i$ are the vector of $i$th particle's position, velocity, angular velocity and temperature, 
$\vec{r}_{i j}=\vec{r}_{i}-\vec{r}_{j}$. The magnitude of the vector ${r}_{i j}=|\vec{r}_{i j}|$ is the distance between particle $j$ and particle $i$. The factor $\lambda_{i j}$ in angular momentum equation is a weighted parameter to account for the different contributions from the particles with different size for the conservation of the angular momentum \cite{pryamitsyn2005coarse}, which is defined as:
\begin{equation}\label{equ:lmd}
\lambda_{i j}=\frac{R_{i}}{R_{i}+R_{j}}=\frac{1}{1+\frac{R_{j}}{R_{i}}},
\end{equation}
where $R_{i}$ and $R_{j}$ denote the radii of particle $i$ and particle $j$. 
If the size of small (solvent and polymer) particle $j$ is far smaller than  NP $i$, then $R_{j} / R_{i}\sim 0$, yields $\lambda_{i j}=1$ and $\lambda_{j i}=0$, thus the angular momentum equation of solvent and polymer in equation (\ref{equ:AM}) vanishes.
$\vec{F}_{i}$ is the total force acting on the particle $i$. $\vec{F}_{i j}$ is the force exerted on particle $i$ by particle $j$, which consists of three parts:
\begin{equation}\label{equ:TF}
\vec{F}_{i j}=\vec{F}_{i j}^{C}+\vec{F}_{i j}^{D}+\vec{F}_{i j}^{R}.
\end{equation}
The conservative force is given by
\begin{equation}\label{equ:FC}
{\vec{F}}_{ij}^C=a_{ij}w_C(r_{ij}){\vec{e}}_{ij}
\end{equation}
In traditional DPD, $a_{ij}$ is the maximum of conservative force,  ${\vec{e}}_{ij}=\vec{r}_{ij}/{r}_{ij}$ is a unit vector,
and the weight function is given by
\begin{equation}\label{equ:WC}
w_C(r_{ij})=
\left\{
\begin{aligned}
&1-\frac{r_{ij}}{R_{CC}}&,\ r_{ij}<R_{CC}
\\
&0						  &,\ r_{ij}> R_{CC}
\end{aligned}
\right.,
\end{equation}
where $R_{CC}$ is the cut-off radius of conservative force.
The dissipative force consists of three parts $\vec{F}_{i j}^{D}=\vec{F}_{i j}^{Dc}+\vec{F}_{i j}^{Ds}+\vec{F}_{i j}^{Dr}$:
(1) The central dissipative force
\begin{equation}\label{equ:FDc}
{\vec{F}}_{ij}^{Dc}=-\gamma_{ij}^cw_D(r_{ij})({\vec{e}}_{ij}\cdot{\vec{v}}_{ij}){\vec{e}}_{ij};
\end{equation}
(2) The shearing dissipative force
\begin{equation}\label{equ:FDs}
{\vec{F}}_{ij}^{Ds}=-\gamma_{ij}^sw_D^2(r_{ij})[{\vec{v}}_{ij}-{(\vec{e}}_{ij}\cdot{\vec{v}}_{ij}){\vec{e}}_{ij}];
\end{equation}
(3) The rotational dissipative force
\begin{equation}\label{equ:FDr}
{\vec{F}}_{ij}^{Dr}=-\gamma_{ij}^sw_D(r_{ij})[{\vec{r}}_{ij}\times(\lambda_{ij}{\vec{\omega}}_i+\lambda_{ji}{\vec{\omega}}_i)],
\end{equation}
where ${\vec{v}}_{ij}={\vec{v}}_i-{\vec{v}}_j$ is relative velocity and $w_D(r_{ij})$ is the weight function for dissipative force.
$\gamma_{ij}^c$ and $\gamma_{ij}^s$ are the central and shearing dissipative coefficients. The rotational dissipative force has the same dissipative parameter $\gamma_{ij}^s$ with the shearing translational dissipative. For $\gamma_{ij}^s=0$, the system recovers to traditional DPD.

The random force is defined by
\begin{equation}\label{equ:FR}
{\vec{F}}_{ij}^Rdt=w_R(r_{ij})[\frac{1}{\sqrt3}\sigma_{ij}^c tr[dW_{ij}]E+\sqrt2\sigma_{ij}^sdW_{ij}^A]\cdot {\vec{e}}_{ij}
\end{equation}
where $dt$ is the time step and $tr[dW_{ij}]$ is the trace of symmetric independent Wiener increment matrix $dW_{ij}$, $E$ is unit matrix.
Anti-symmetric part of $dW_{ij}$ is $dW^A_{ij}$.
$\sigma_{ij}^c$ and $\sigma_{ij}^s$ are the parameters in random force and $w_R(r_{ij})$ is the weight function of random force, which  follows the fluctuation-dissipative theorem
\begin{equation}\label{equ:FDT1}
(\sigma_{ij}^c)^2=\frac{4k_B\gamma_{ij}^c\Theta_i\Theta_j}{\Theta_i+\Theta_j} , 
(\sigma_{ij}^s)^2=\frac{4k_B\gamma_{ij}^s\Theta_i\Theta_j}{\Theta_i+\Theta_j}, 
\end{equation}  
\begin{equation}\label{equ:FDT2}
w_D(r_{ij})=w^2_R(r_{ij}))=(1-\frac{r_{ij}}{R_{CD}})^{s},
\end{equation}  
where $k_B$ is the Boltzmann constant,
$R_{CD}$ is the cut-off radius for dissipative and random force,
and $s$ is the exponent of weight function\cite{Fan2006Simulating}.

$\vec{q}_{i}$ is the total heat flux that acts on the particle $i$. $\vec{q}_{i j}$ is the heat flux exerted on particle $i$ by particle $j$, which consists of three parts:
\begin{equation}\label{equ:qij3Part}
q_{ij}=q^{C}_{ij}+q^{V}_{ij}+q^{R}_{ij}
\end{equation}
in which $q_{ij}^C$ and $q^{R}_{ij}$ is the collisional and random heat flux, given by
\begin{equation}\label{equ:qc}
\begin{aligned}
&q^{C}_{ij}=k_{ij}w_{CT}(r_{ij})(\frac{1}{\Theta_i}-\frac{1}{\Theta_j}),\\
&q_{ij}^{R}dt=\beta_{ij}w_{RT}(r_{ij})dW_{ij}^e.\\ 
\end{aligned}
\end{equation}
where $k_{ij}=C_{vi}C_{vj}\kappa\Theta_{ij}^2/k_B$ in which $\kappa$ is the mesoscale heat friction coefficient. $\beta_{ij}^2=k_{ij}k_B$.
The weight function $w_{CT}(r_{ij})$ and $w_{RT}(r_{ij})$ are given as $w_{CT}(r_{ij})=w_{RT}^2(r_{ij})=1-r_{ij}/R_{CT}$ with the cut-off radius $R_{CT}$.

The viscous heat flux contain three parts $q^{V}_{ij}=q^{Vc}_{ij}+q^{Vs}_{ij}+q^{Vr}_{ij}$:
(1) viscous heat flux for central force
\begin{equation}\label{equ:qvc}
\begin{aligned}
  &(C_{vi}+C_{vj}) q^{Vc}_{ij}dt
  = w_D(r_{ij})[(\vec{e_{ij}} \cdot \vec{v_{ij}})^2-\frac{\sigma^c_{ij}}{m}]dt \\
  &-\sigma^c_{ij}w_R(r_{ij})(\vec{e_{ij}} \cdot \vec{v_{ij}})\frac{tr[dW_{ij}]}{3}  \\
\end{aligned}
\end{equation}
(2) viscous heat flux for shearing force
\begin{equation}
\begin{aligned}
&(C_{vi}+C_{vj}) q_{i j}^{V s}dt  =\\
& w_D(r_{i j})[-\gamma_{i j}^s(\vec{e_{ij}} \cdot \vec{v_{ij}})^2+\gamma_{i j}^s(\vec{v_{ij}} \cdot \vec{v_{ij}})-\frac{2(\sigma_{ij}^s)^2}{m}]dt \\
&-\sigma_{i j}^s w_R(r_{i j}) d W_{i j}^A \cdot \vec{e_{ij}} \cdot \vec{v_{ij}}\\
\end{aligned}
\end{equation}
(3) viscous heat flux for rotational force
\begin{equation}\label{equ:VHF_RF}
\begin{aligned}
(C_{vi}+C_{vj}) q_{ij}^{Vr}=
\gamma_{ij}^sw_D(r_{ij})[{\vec{r}}_{ij}\times(\lambda_{ij}{\vec{\omega}}_i+\lambda_{ji}{\vec{\omega}}_j)] \cdot \vec{v_{ij}}
\end{aligned}
\end{equation}

In this paper, we study the thermophoresis problem of nanoparticles, where the rotation of the particles is not a dominant factor. Therefore, we do not calculate Eq. (\ref{equ:AM}). As a result, the rotational dissipative force (Eq. (\ref{equ:FDr})) and the viscous heat flux for rotational force (Eq. (\ref{equ:VHF_RF})) are also not accounted for in the simulation.

\subsection{Mapping of units}
In seDPD, the energy is conservative, the summation of potential energy, kinetic energy and internal energy is a constant
In equilibrium, potential energy is minimized, the energy conserving equation is
\begin{equation}
\begin{aligned}
    &dE_{k}^{*}=dE_{I}^{*}\\
    &m^{*}v^{*}dv^{*}=\frac{m^{*}}{\rho^{*}}C_{v}^{*}d\Theta^{*}\\
\end{aligned}
\end{equation}
Using the following dimensionless variables
\begin{equation}\label{equ:Dless}
    v=\frac{v^*T}{L};
    m=\frac{m^{*}}{M};
    \rho=\frac{\rho^{*}L^3}{M};
    \Theta=\frac{k_B^{*}\Theta^{*}}{ML^2/T^2};
\end{equation}
the dimensionless energy conserving equation is written as
\begin{equation}\label{equ:HBFless}
 \frac{1}{C_{v}}vdv=C_{v}d\Theta.
\end{equation}
and $C_{v}$ can be written as\cite{Li2014Energy}:
\begin{equation}\label{equ:Cv}
    C_{v}=\frac{C_{v}^{*}L^3}{\rho k_B^{*}};
\end{equation}
in which is $k_B^{*}$ is the Boltzmann constant, $C_{v}^{*}$ is the volumetric heat capacity of fluid.
The equation (\ref{equ:Cv}) indicate that the large length scale will lead to a large dimensionless heat capacity $C_v$. 

For water at $300k$, $k_B^{*}=1.381 \times 10^{-23}JK^{-1} $, $C_{v}^{*}=4.167 \times 10^{6} Jm^{-3}K^{-1}$.
when we choose the length unit $L=100nm$, temperature of reference $300K$ and number density $\rho=3.0$, we get the dimensionless volumetric heat capacity $C_{v}=1\times 10^8$.

\subsection{Method to measure Soret force}
Considering a particle driven by the force proportional to the temperature gradients $F^T=-S_T\nabla \Theta$ and subject to the stochastic force.
The coordinate of Brownian particle $x$ is governed by an over-damped non-linear Langevin equations
\begin{equation}\label{equ:NLEsoret}
    dx=-D_0S_T\nabla \Theta dt+\sqrt{2D_0}d\mathcal{W}
\end{equation}
in which $\mathcal{W}$ is the wiener process, $D_0$ is the diffusivity for particle diffusion in simple fluid.
The unsteady state method to measure Soret force is to calculate the mean displacement of particles $\langle x\rangle/t=-D_0S_T \nabla \Theta$.

A method to calculated Soret force in steady state is proposed by Duhr {\it et al.}\cite{Duhr2006Thermophoretic}.
The corresponding Fokker-Planck function of equation (\ref{equ:NLEsoret}) is given by
\begin{equation}\label{equ:FPsoret}
\left\{\begin{aligned}
&\frac{\partial \rho(x,t)}{\partial t}=-\frac{\partial  J_T+J_D}{\partial x} \\
&J_T=-D_0  S_T\nabla \Theta  \rho(x,t), 
\\
&J_D=-D_0  \frac{\partial }{\partial x}  \rho(x,t), 
\\
\end{aligned}\right.
\end{equation}
where $J_T$ and $J_D$ is the mass flux associated with a thermal gradient and concentration gradient.
In steady state, the total mass flux is zero ($J_T=-J_D$), yielding
\begin{equation} \label{equ:rhoexp}
    d\rho/\rho=-S_Td\Theta
\end{equation}
which gives a convincing method to obtain the Soret coefficient $S_T=-d{\log(\rho)}/dT$ from the experimental or numerical results of the concentration and temperature in steady state.

\section{Results and discussion}\label{sec3}
\subsection{Validation of seDPD}
The algorithmic validity of seDPD for simulating thermophoresis is examined by considering the effects of fluid viscosity, thermal conductivity, and hydrodynamic drag force on nanoparticles.

\begin{figure}[h]
\centering
\includegraphics[width=0.48\textwidth]{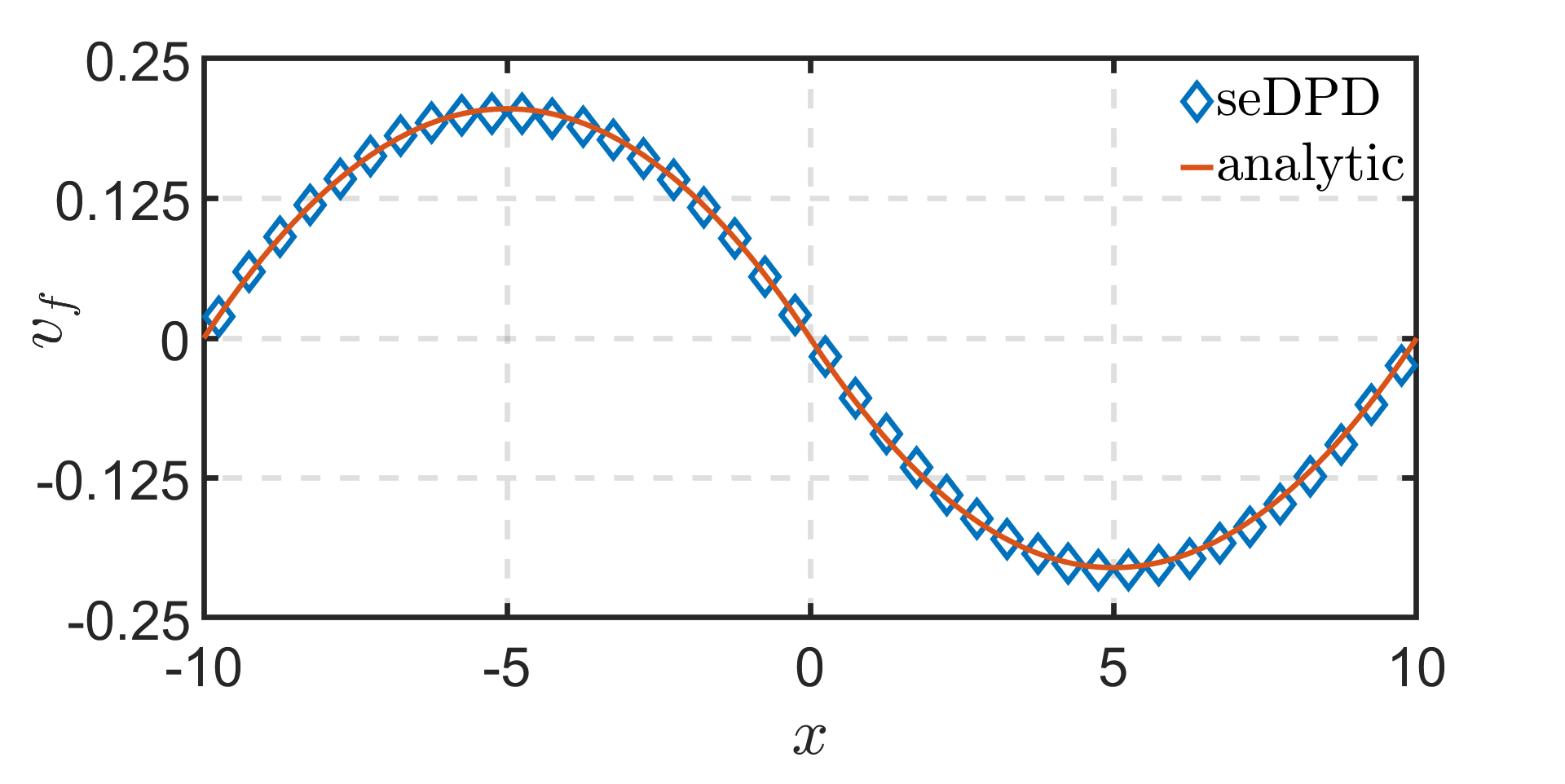}
  \captionsetup {justification=raggedright,singlelinecheck=false}
	\caption{The velocity profile for the Poiseuille flow in seDPD. The parameters is set to $\rho=3.0$, $a_{ij}=25$, $\gamma_c=4.5$, $\gamma_s=0$, $s=0.4$, $R_{CC}=R_{CD}=1.0$.
 The driven force is $F=0.01$.
}
	\label{img:poiFlow}
\end{figure}

First, we simulate the Poiseuille flow to validate the hydrodynamic properties of the seDPD method. We implement the non-slip boundary condition using reverse Poiseuille flow, which was proposed by Backer {\it et al.}\cite{backer2005poiseuille} and is widely used in DPD simulation\cite{Li2014Energy,Zhang2022eDPDm}.
The seDPD parameters for the Poiseuille flow are listed in the caption of Fig. \ref{img:poiFlow}.
The simulation domain is $20 \times 10 \times 8$ with DPD units.
We apply two equal and opposite driving forces of magnitude $F=0.01$ along the $y-$direction.
The velocity profile ($x \in [0,10]$) is shown in Fig. \ref{img:poiFlow}, which is in good agreement with the analytical solution of the Navier-Stokes equations.
\begin{equation}
    v_f=\frac{\rho Fx}{2\eta_s}({L_x}/{2}- \left| x \right|),
\end{equation}
where half of the simulation domain in $x-$direction is $L_x/2=10$, and the dynamic viscosity $\eta_s=1.83$, which agrees with the results $1.8368$ in our previous study\cite{Lu2021potential} for sDPD with the same parameters.

\begin{figure}[h]
	\centering
	\includegraphics[width=0.48\textwidth]{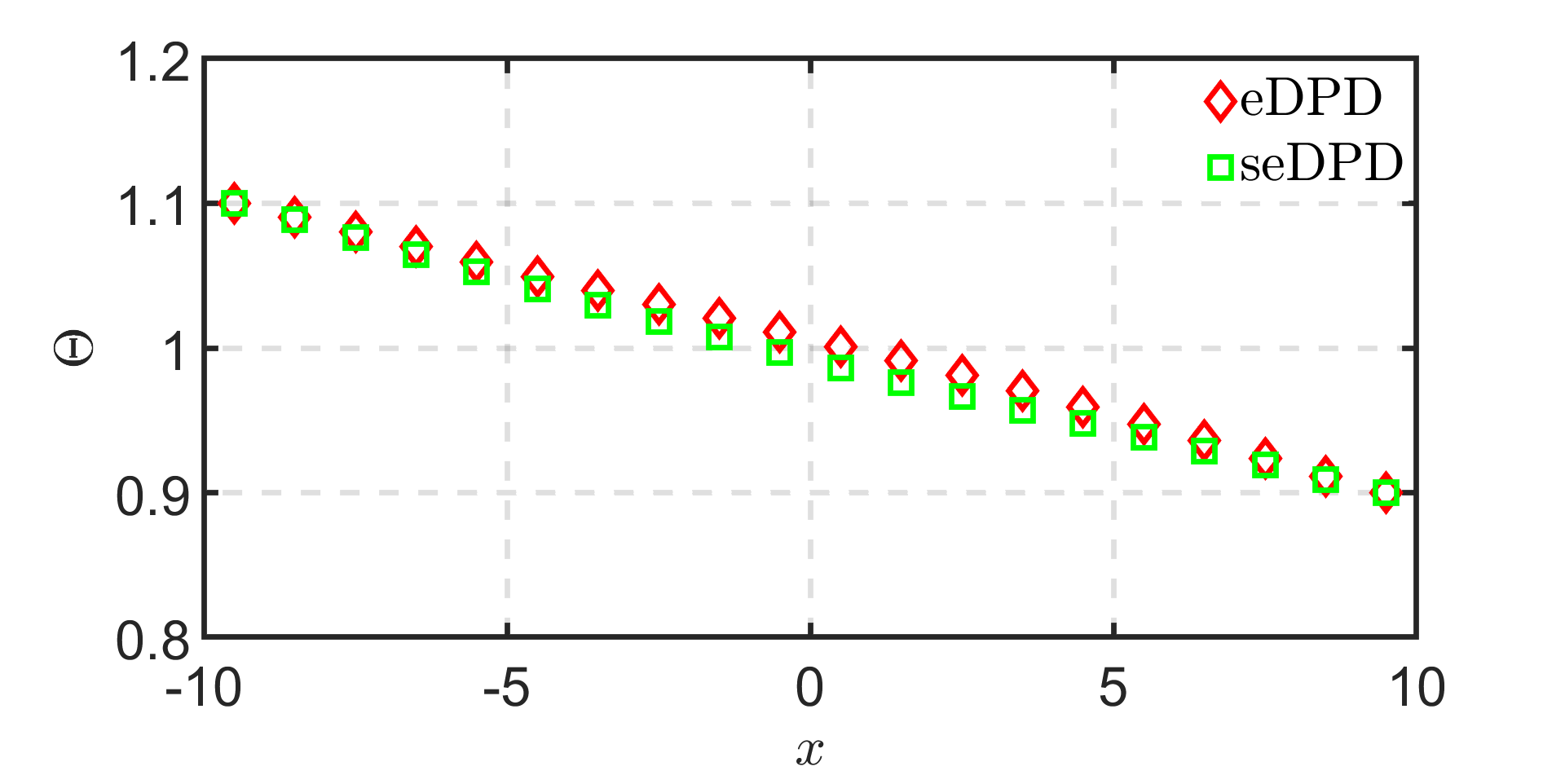}
\captionsetup {justification=raggedright,singlelinecheck=false}
\caption{The temperature profile for seDPD and eDPD. The parameters are set to $\kappa=10^{-8}$, $C_v=10^{8}$, $R_{CT}=1.0$.
}
	\label{img:T_profile}
\end{figure}

\begin{figure}[h]
	\centering
	\includegraphics[width=0.48\textwidth]{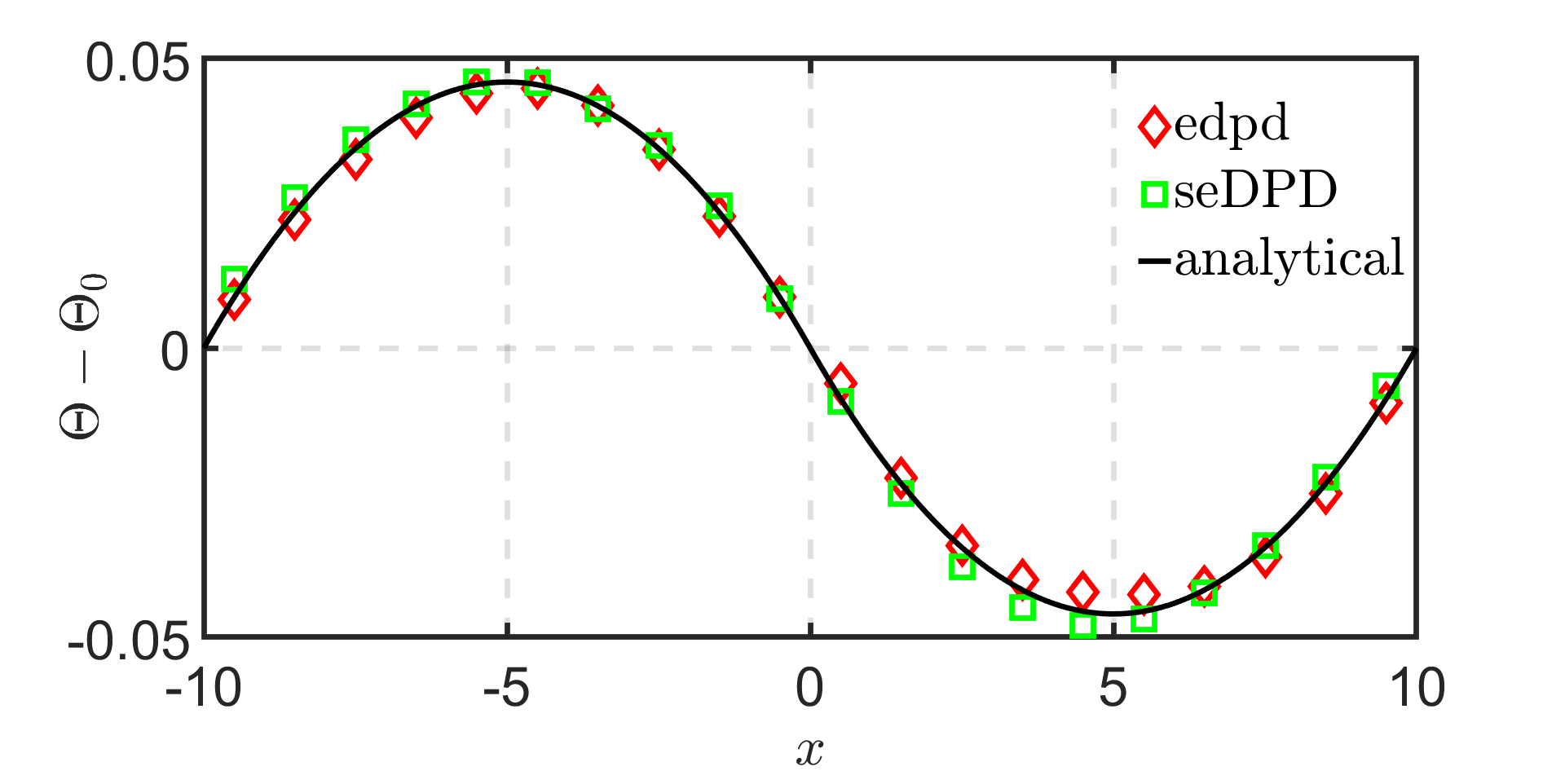}
\captionsetup {justification=raggedright,singlelinecheck=false}
\caption{The temperature profile for seDPD and eDPD. The parameters are set to $\kappa=10^{-8}$, $C_v=10^{8}$, $R_{CT}=1.0$.
}
	\label{img:HRPF_sedpd_edpd}
\end{figure}

A steady state heat conduction is simulated to validate the thermal conduction in seDPD. The stationary fluid is confined between the hot wall at $T_H=1.1$ and the cold wall at $T_c=0.9$. The solid wall is constructed by freeze the fluid particles. The fluid particle penetrates the solid wall and bounces back to the flow field. As shown in Fig. \ref{img:T_profile}, the temperature $\Theta$ is linear with position, which is the steady solution for the heat transfer equation.
The thermal conductivity of the fluid is obtained by simulating the heat conduction analog of reverse Poiseuille flow,  which was proposed by Li {\it et al.}\cite{Li2014Energy}. 
Using the same simulation box of the RPF, one-half of the fluid domain is heated, and the other-half is cooled.
As shown in Fig. \ref{img:HRPF_sedpd_edpd}, the temperature profile ($x\in [0 ,10]$) agrees well with the analytical solution of the heat conduction equation.
\begin{equation}
    \Theta-\Theta_0=\frac{\rho Qx}{2\lambda_s}({L_x}/{2}-\left| x \right|),
\end{equation}
where $\Theta_0$ is the temperature of the seDPD system before heating/cooling, and $\lambda_s$ is the thermal diffusivity.
As shown in Fig. \ref{img:T_profile} and Fig. \ref{img:HRPF_sedpd_edpd}, the temperature profile in seDPD agrees with those of the eDPD model, indicating that we can use the seDPD model to obtain the same thermal properties as eDPD.

\begin{figure}[h]
	\centering
	\includegraphics[width=0.48\textwidth]{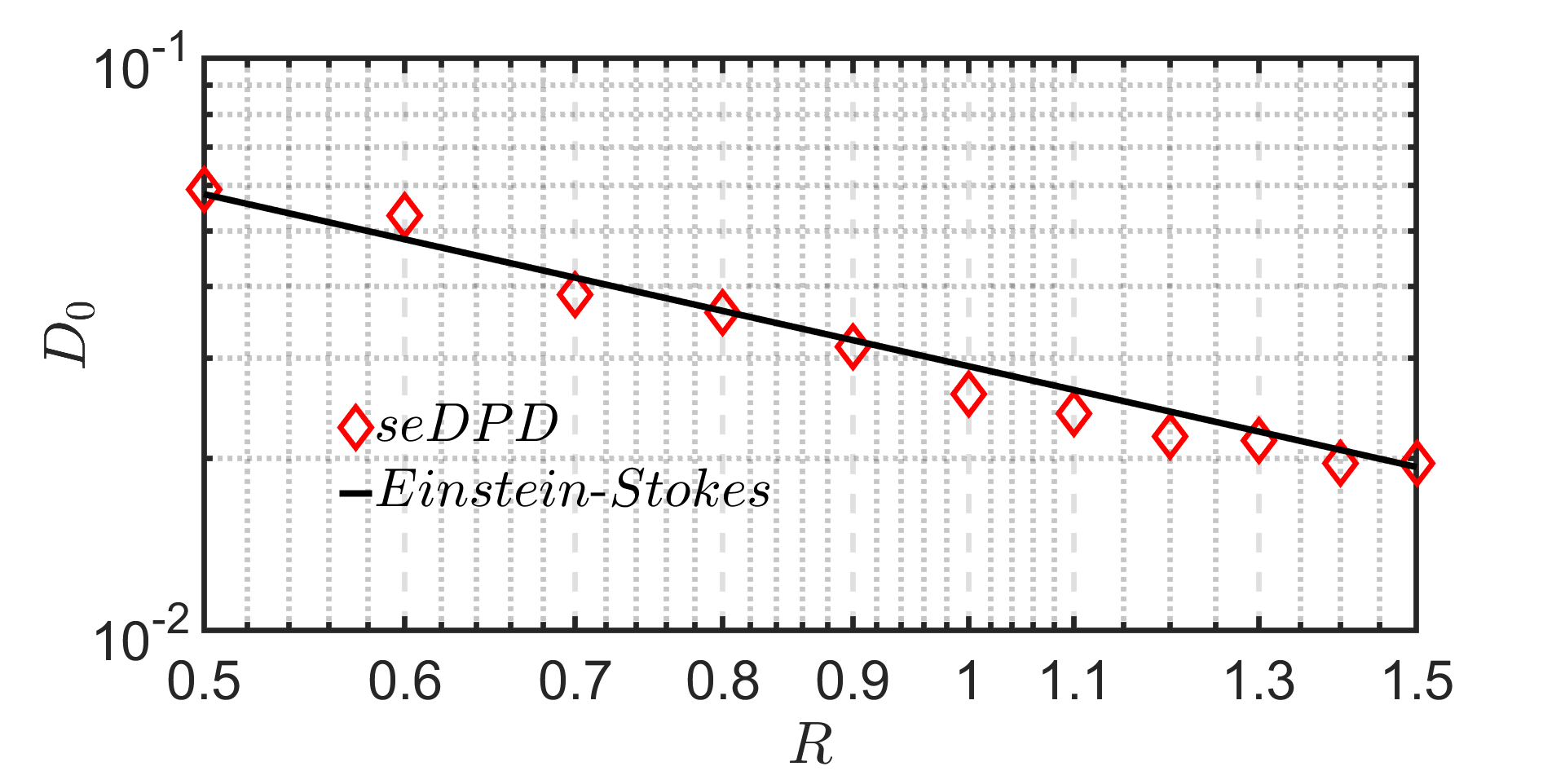}
\captionsetup {justification=raggedright,singlelinecheck=false}
\caption{The free diffusivity of nanoparticles in simple fluid.
}
	\label{img:D0_seDPD}
\end{figure}

Nanoparticles moving in a fluid under a temperature gradient experience a Soret force that balances the hydrodynamic drag force, i.e., $F^T=\gamma v$, where $\gamma$ is the friction coefficient.
It is crucial to verify that the nanoparticles receive the correct hydrodynamic drag for studying thermophoresis. The friction coefficient can be calculated from the diffusivity of the nanoparticles using Einstein's formula $D_0=k_B\Theta/\gamma$.
The friction coefficient should follow the Stokes formula for the hydrodynamic drag of nanoparticles, i.e., $\gamma=6\pi R \eta_s$.
As shown in the Fig. \ref{img:D0_seDPD}, the self-diffusion rate of nanoparticles in the fluid is inversely proportional to the size of nanoparticles and follows the Einstein-Stokes formula.
This demonstrates that the seDPD method can accurately capture the hydrodynamic drag of nanoparticles in the fluid.

\subsection{Soret effect in seDPD}
To model the Soret effect in a DPD system, the key problem is to apply $F^T$ on a nanoparticle with a finite size $R$. A modified conservative force is utilized for NP and solvent:
\begin{equation}\label{equ:SoretForce}
\begin{aligned}
   &{\vec{F}}_{ij}^{C}=a_{ij}w_C(r_{ij}){\vec{e}}_{ij}+{\vec{F}}_{ij}^{CT} ,\\
   &{\vec{F}}_{ij}^{CT}=-a^T_{ij}(\Theta_i-\Theta_j)w_{CT}(r_{ij})  \vec{e_{ij}}.\\
\end{aligned}
\end{equation}
To satisfy Newton's third law ($F_{ij}^{CT}=-F_{ji}^{CT}$), the parameter must be set to $a_{ij}^{T}=-a_{ji}^T$. For a positive Soret coefficient of the particle $i$, $a^T_{ij}$ should be positive, and when $T_{i}>T_{j}$ the particle with higher temperature will move to the region with lower temperature.
The weight function $w_C$ in Eqs. (\ref{equ:SoretForce}) for the interaction of NP and solvent has an exponential form\cite{pan2009rheology,afrouzi2018systematic}: 
\begin{equation}\label{equ:WCexp}
w_C\left(r_{ij}\right)=
\left\{
\begin{aligned}
&\frac{1}{1-e^{-b_{ij}}}\left(e^{-\frac{b_{ij}r_{ij}}{R_{}+\delta}}-e^{-b_{ij}}\right) &, \ r_{ij}<R_{}+\delta
\\
&0  &, \ r_{ij}> R_{}+\delta 
\end{aligned}
\right.,
\end{equation}
where $b_{ij}$ is the factor of weight function, $R$ is the 
hydrodynamic radius of Brownian particles, and $\delta$ is the effective radius of fluid particles\cite{wang2018parametric}.
By applying the exponential form weight function, the fluid particle cannot penetrate into the NP.

\begin{figure}[h]
	\centering
    \captionsetup {justification=raggedright,singlelinecheck=false}
	\includegraphics[width=0.48\textwidth]{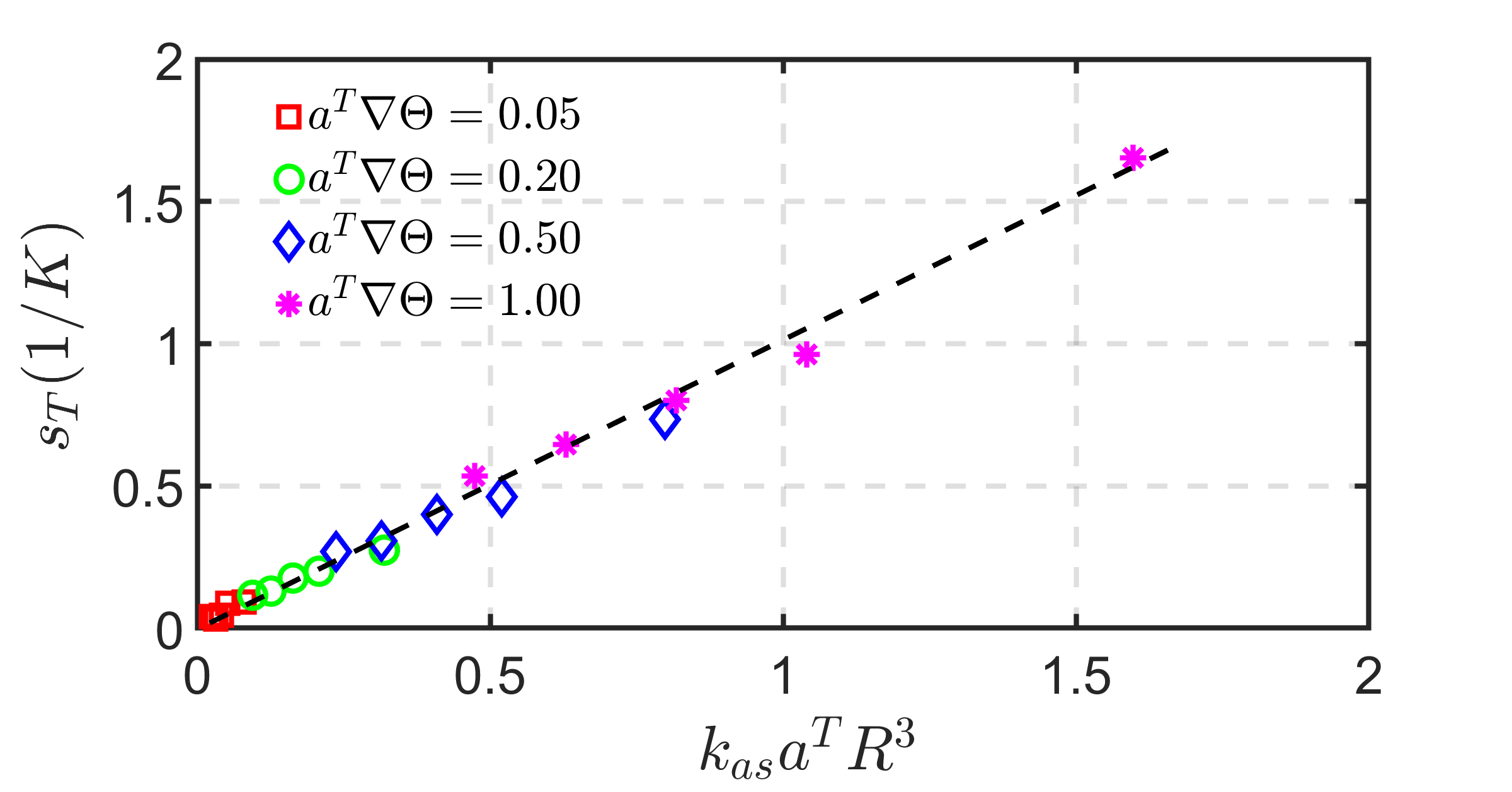}
	\caption{The numerical results of Soret coefficient $S_T$ is linear with $a^TR^3$. $k_{as}$ is the prefactor.
}

	\label{img:sT_vs_aT_R}
\end{figure}

The modified conservative force (Eqs. \ref{equ:SoretForce}) provides the in-homogeneous interaction between nanoparticles and fluid particles in the temperature gradient, which results in the thermophoretic force on the nanoparticles parallel to the temperature gradient.
As shown in Fig. \ref{img:sT_vs_aT_R}, we obtain the following scaling law for the Soret coefficient and Soret force:
\begin{equation}\label{equ:SoretvsaT}
\begin{aligned}
&S_T \sim a^TR^3, \\
&F^T \sim a^T\nabla \Theta R^3, \\
\end{aligned}
\end{equation}
A previous experiment\cite{Duhr2006Thermophoretic} showed that the Soret coefficient $S_T$ grows linearly with the square of the particle size.
Therefore, we set the parameter $a^T$ to $a^T = \alpha^T R^{-1}$, where $\alpha^T$ is a physical parameter that does not depend on the particle size.

\subsection{Nanoparticle driven by Soret force transport in a polymer network}

\begin{figure}[h]
	\centering
    \captionsetup {justification=raggedright,singlelinecheck=false}
	\includegraphics[width=0.49\textwidth]{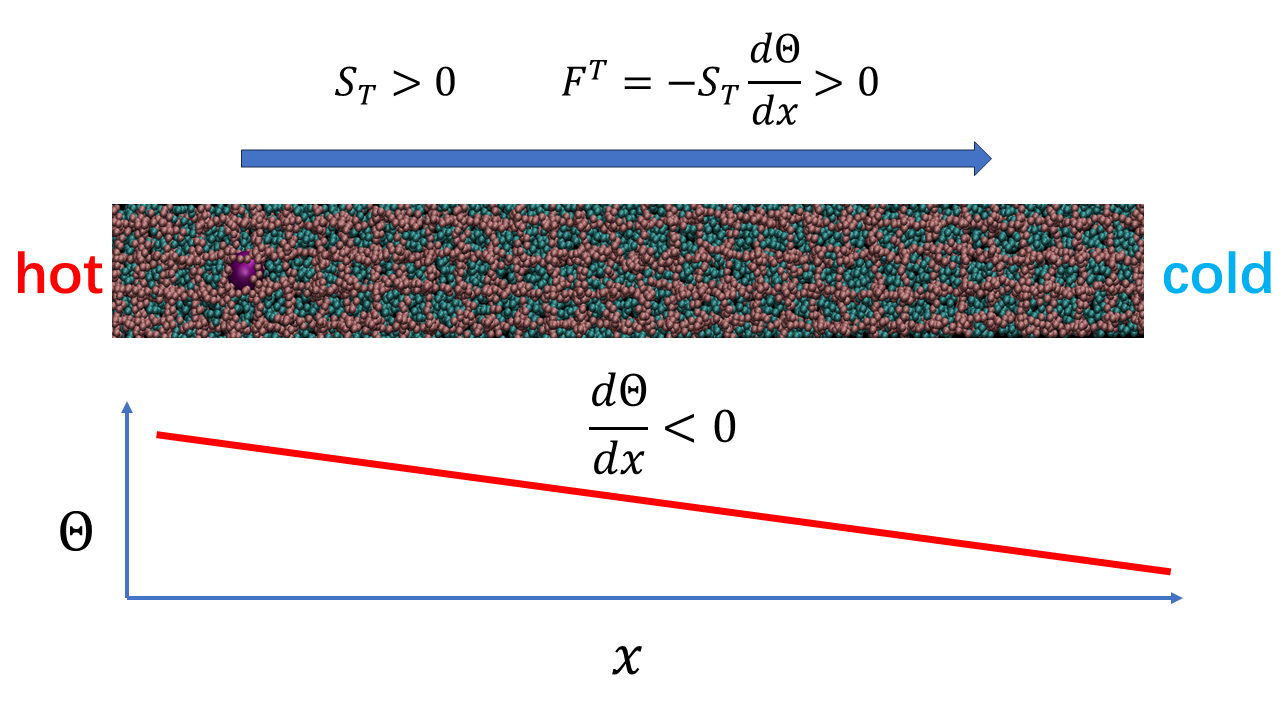}
	\caption{Nanoparticle (purple) driven by the Soret force moving in the polymer network (red). The fluied particles (blue) are filled to keep the number density $\rho=3$.
}

	\label{img:sche}
\end{figure}

By applying modified conservative force (Eqs. (\ref{equ:SoretForce})), the Soret force can be successfully applied on the nanoparticle in simple fluid. 
As a physical problem of great value for fundamental and applied research, the motion of the nanoparticle driven by the Soret force in the polymer network is worth analyzing.
As shown in Fig. \ref{img:sche}, the polymer network with periodic structure is immersed in the fluid with a temperature gradient. The nanoparticle embedded in the polymer network driven by the Soret force.
When the Soret coefficient is positive $S_T>0$, the Soret force $F^T$ is opposite to the direction of temperature gradient, thus the nanoparticle move from the hot boundary to the cold boundary. 

In the polymer network, there are $N=8$ DPD particles between every two neighbouring junctions, and the average distance between two neighbouring junctions is mesh size $a_x=2.33$.
The adjacent DPD particles are connected by harmonic spring
\begin{equation}\label{equ:EBD}
	E_{b}(r_{ij})=K_b(\frac{r_{ij}}{l_b}-1)^2,
\end{equation}
in which $K_b=125$ stands for the energy of the bond, $l_b=0.45$ is the equilibrium distance between two DPD particles, thus the counter length of polymer stand is $L_c=(N-1)l_b=3.15$.

The simulation coefficients are shown in Table \ref{tb:para}, where the first column indicates the type of particle: $w$ for water, $n$ for network, and $p$ for nanoparticle.
Due to the exponential conservative force (Eqs. (\ref{equ:WCexp})), no DPD particles exist within the radius $R$ of the nanoparticle, and only a spherical shell with a thickness of $\delta$ contains particles that interact directly with the nanoparticle.
According to our previous work\cite{Lu2021potential}, the cutoff radius of the dissipative force $R_{CD}$ needs to be larger than that of the conservative force $R_{CC}$ for the interactions between nanoparticles and polymers as well as nanoparticles and fluids.
Similarly, we chose a larger cutoff radius of the heat flux $R_{CT}$ than that of the conservative force, to ensure effective heat transfer and Soret force between the nanoparticles and the fluid and network particles.
We also selected a large thermal friction $\kappa$ coefficient to maintain the temperature gradient undisturbed by the motion of the nanoparticles, and to ensure the thermal equilibrium of the nanoparticles with the local temperature. 
Volumetric heat capacity $C_v=1 \times 10^8$ was determined by Eqs. (\ref{equ:Cv}), using a characteristic size of $L=100nm$, a number density of $\rho=3$, and the physical parameters of water at $300 K$.

\begin{table}[t]
\caption{\label{tb:para}
The simulation coefficients
}
\begin{ruledtabular}
\begin{tabular}{ccccccccccc}
\textrm{pair}&
\textrm{$a_{ij}$}&
\textrm{$\gamma^c$ }&
\textrm{$\gamma^s$ }&
\textrm{$s$}&
\textrm{$R_{CC}$}&
\textrm{$R_{CD}$}&
\textrm{$\kappa$ }&
\textrm{$C_v$ }&
\textrm{$R_{CT}$ }&
\textrm{$a^T$ }
\\
\colrule
ww/nn/wn & 25 & 4.5& 0 & 0.4 &1.0&1.0&$10^{-8}$&$10^{8}$&1.0&0\\
wp/np & Ref. \cite{Lu2021potential} & 3.5& 5.5 & 0.4 & $R$ & Ref. \cite{Lu2021potential} &$10^{-8}$&$10^{8}$& $R+R_s$ &Eqs.\ref{equ:SoretvsaT}\\
\end{tabular}
\end{ruledtabular}
\end{table}

We studied the transport characteristics of nanoparticles in polymer network with different sizes ($R=1.3,1.4,1.5$) under different thermophoretic force.
When the size of a nanoparticle is larger than the mesh size of the polymer network $2R>a_x$, the diffusion of the nanoparticle is hop-dominated\cite{cai2015hopping}.
Thus, the local confinement that the polymer network exerts on the nanoparticle can be described by a periodic potential energy\cite{Lu2022Double,Lu2021potential,Dell2013Theory,cai2015hopping}.
The size of the nanoparticles is also chosen to be smaller than the contour length of the strands in the polymer network $2R<L_c$, thus the potential energy is dominated by entropy.
When driven by a constant force, the nanoparticle diffusion occurs in a tilted periodic potential, also referred to the washboard potential (WBP).
In the present work, the driving force is realized by the Soret effect.

From the trajectories of the nanoparticle $x_n(t)$, the travel distance is $\langle x_n(t) \rangle$ and the displacement variance is $\delta \langle x_n(t) \rangle=\langle x_n^2(t) \rangle-\langle x_n(t) \rangle^2$ can be calculated using the ensemble average.
The transport characteristics of foremost interest is the drift velocity and diffusivity in long time stage
\begin{equation}\label{equ:vL}
\left\{
\begin{aligned}
v_L &=\lim_{t\rightarrow \infty} v(t)=\lim_{t\rightarrow \infty} \langle x_n(t) \rangle/t, \\
D_L &=\lim_{t\rightarrow \infty}D(t)=\lim_{t\rightarrow \infty}\frac{\delta\langle x^2(t) \rangle}{2t}. \\
\end{aligned}\right.
\end{equation}

\begin{figure}[h]
	\centering
 \includegraphics[width=0.48\textwidth]{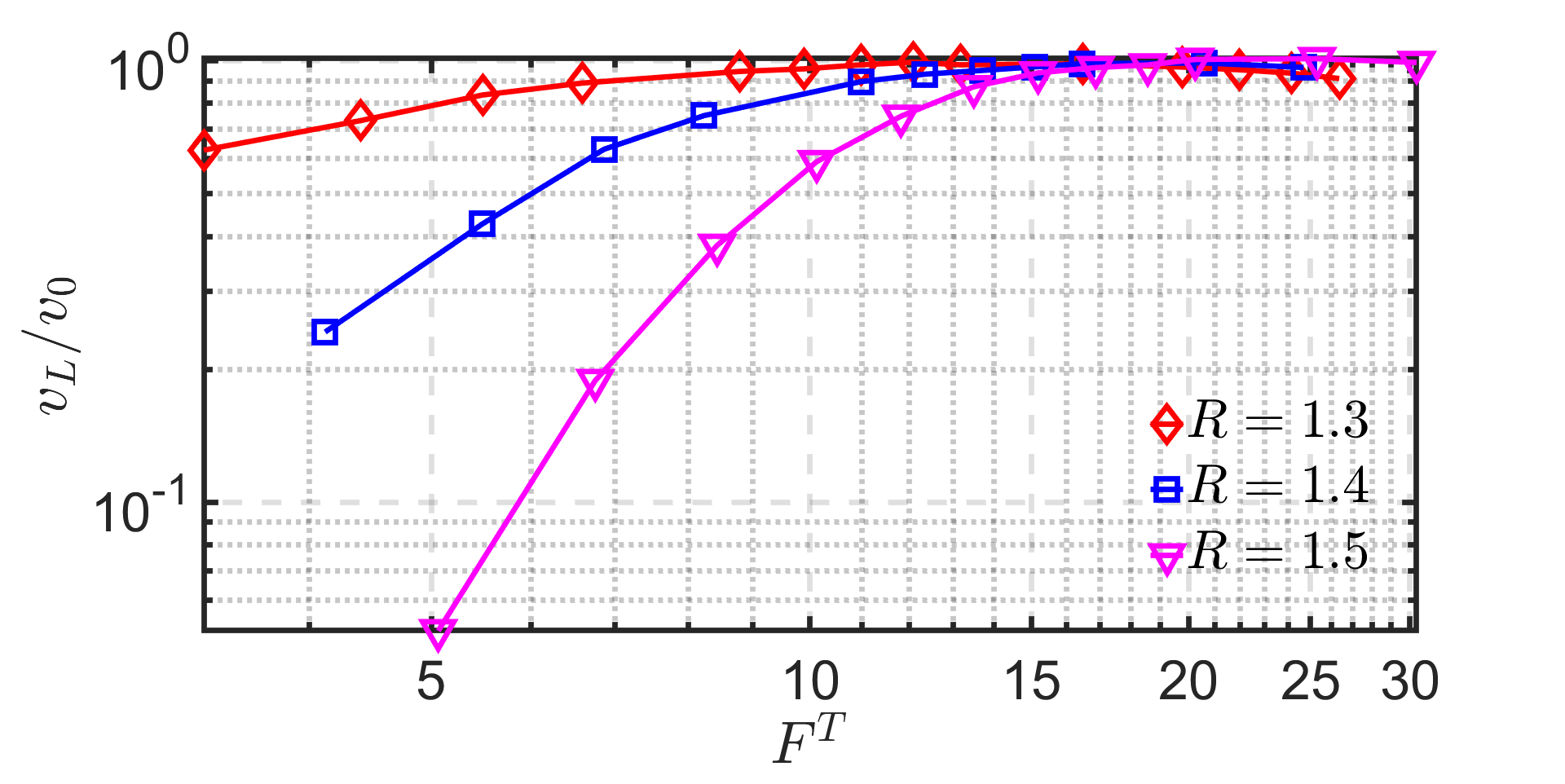}
\captionsetup {justification=raggedright,singlelinecheck=false}
\caption{Drift velocity of nanoparticles in polymer network is converged to that in simple fluid, as the Soret force increase.
}
	\label{img:VL_V0_R13to15}
\end{figure}

As shown in Fig. \ref{img:VL_V0_R13to15}, the normalized drift velocity of nanoparticles confined in polymer networks, $v_L/v_0$, is plotted as a function of the Soret force, $F^T\sim a^T\nabla\Theta R^3$. Here, $v_0$ is the drift velocity of nanoparticles without confinement, which is given by
$$
v_0=D_0F^T\sim D_0 a^T\nabla\Theta R^3
$$
The drift velocity of confined nanoparticles $v_L$ increases gradually and approaches the free drift velocity $v_0$ as the Soret force $F^T$ increases. When the Soret force is small, the confinement effect lowers the velocity of nanoparticles compared to the unconfined case, that is $v_L<v_0$. 
This results is qualitatively consistent with previous studies on the diffusion of Brownian particles in a tilted periodic potential $U$\cite{Reimann2001Giant,Reimann2008Weak,kim2017giant}.

The confinement effect is determined by the height of the potential barrier, which depends on the ratio of the nanoparticle and mesh size\cite{Lu2021potential,cai2015hopping}, that is, the confinement parameters $2R/a_x$. Therefore, the velocity reduction is more significant for larger nanoparticles, due to the higher potential barrier.

\begin{figure}[h]
	\centering
	\includegraphics[width=0.48\textwidth]{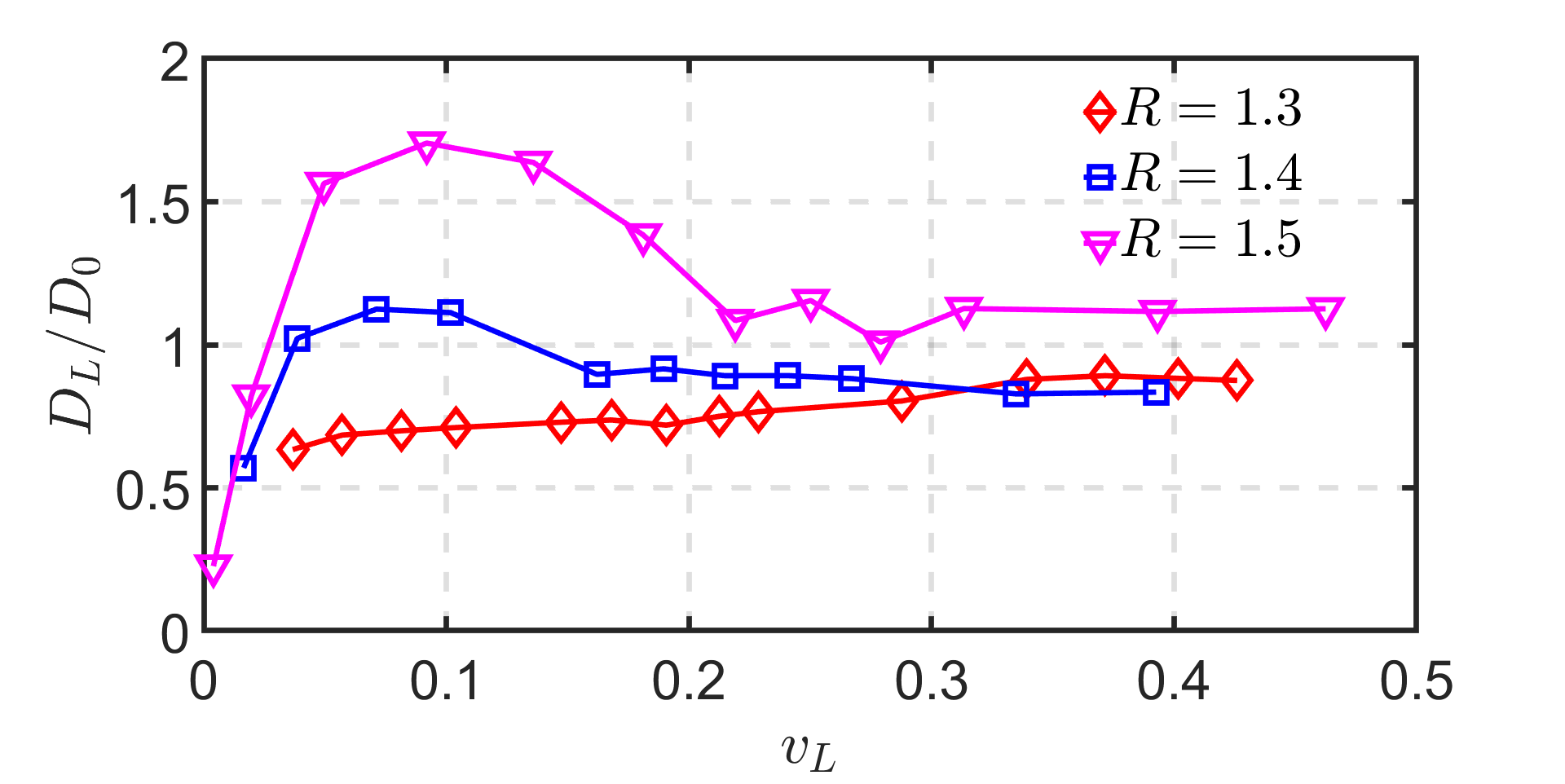}
\captionsetup {justification=raggedright,singlelinecheck=false}
\caption{Variation of normalized diffusivity $D_L/D_0$ with drift velocity $v_L$.
}
	\label{img:DL_R13to15}
\end{figure}

As shown in Fig. \ref{img:DL_R13to15}, the normalized diffusivity $D_L/D_0$ of nanoparticle diffusion in the polymer network is plotted against the drift velocity $v_L$, which monotonically increases with the Soret force $F^T$.
The diffusion of nanoparticles in polymer networks exhibits a non-monotonic behavior as the function $v_L$, except for the nanoparticles with size $R=1.3$, that is, the diffusion coefficient reaches a peak value that is larger than the free diffusion coefficient at a critical thermophoretic force. 
This phenomenon is known as giant acceleration of diffusion (GAD)\cite{Reimann2001Giant,Reimann2008Weak,kim2017giant}, which is a non-equilibrium transport phenomenon in a tilted periodic potential. Since the potential energy exerted by the polymer network on the nanoparticles is dominated by entropy, this result numerically confirms that the Brownian particles crossing the entropy barrier can exhibit GAD\cite{Burada2008Entropic,kim2017giant}.
For higher entropy barriers, the enhancement of diffusivity is more pronounced for larger nanoparticles, which is qualitatively consistent with the GAD results in tilted periodic potentials\cite{Reimann2001Giant,Venditti2022Physica}.

\section{Conclusion}\label{sec4}
This paper presents a novel extension of dissipative particle dynamics (DPD), called the single-particle energy-conserving dissipative particle dynamics (seDPD) method, which combines the features of single-particle DPD and energy-conserving DPD to simulate the thermophoresis of nanoparticles with a single DPD particle under temperature gradients. 

The method accounts for the heat flux arising from the non-central dissipative and random forces, ensuring the conservation of kinetic and internal energy. The validity and reliability of the method and algorithm are verified. The thermophoretic force exerted on the nanoparticles is obtained by modifying the conservative force. Parameterization of the thermophoretic force is established on the basis of the DPD parameters and the nanoparticle size. 

The transport of nanoparticles driven by the thermophoretic force in a polymer network is investigated. The phenomenon of giant acceleration of diffusion (GAD) of nanoparticles crossing the entropic barrier is observed. This paper demonstrates a reliable novel DPD method that can simulate systems involving nanoparticles and temperature fields simultaneously and shows numerically that Brownian particles can exhibit GAD when crossing the entropic barrier.

This model is not limited to thermophoresis of nanoparticles and can readily be extended to investigate various system including temperature and nanoparticles.

\begin{acknowledgments}
This research is supported by the Natural Science Foundation
of China (Nos. 12332016, 11832017 and 12172209).
\end{acknowledgments}

\section*{Data Availability}
The data that support the findings of this study are available from the corresponding author upon reasonable request.
\bibliographystyle{unsrt}
\bibliography{aipsamp}

\end{document}